\documentclass[prb,showpacs,twocolumn,superscriptaddress]{revtex4-1}

\usepackage{amssymb}
\usepackage{graphicx}
\usepackage{amsmath,amssymb}
\begin{document}

\title{Compensation effect in carbon nanotube quantum dots coupled to polarized electrodes in the presence of spin-orbit coupling}
\author{Lin Li}
\affiliation{Center of Interdisciplinary Studies and Key Laboratory for Magnetism and Magnetic Materials of the Ministry of Education, Lanzhou University, Lanzhou 730000,
China}
\author{Yang-Yang Ni}
\affiliation{Center of Interdisciplinary Studies and Key Laboratory for Magnetism and Magnetic Materials of the Ministry of Education, Lanzhou University, Lanzhou 730000,
China}
\author{Tie-Feng Fang}
\affiliation{Center of Interdisciplinary Studies and Key Laboratory for Magnetism and Magnetic Materials of the Ministry of Education, Lanzhou University, Lanzhou 730000,
China}
\author{Hong-Gang Luo}
\affiliation{Center of Interdisciplinary Studies and Key Laboratory for Magnetism and Magnetic Materials of the Ministry of Education, Lanzhou University, Lanzhou 730000,
China}
\affiliation{Beijing Computational Science Research Center, Beijing 100084, China}
\date{\today}

\begin{abstract}
We study theoretically the Kondo effect in carbon nanotube quantum dot attached to
polarized electrodes. Since both spin and orbit degrees of freedom
are involved in such a system, the electrode polarization
contains the spin- and orbit-polarizations as well as the Kramers
polarization in the presence of the spin-orbit coupling. In this
paper we focus on the compensation effect of the effective fields
induced by different polarizations by applying magnetic field. The
main results are i) while the effective fields induced by the spin-
and orbit-polarizations remove the degeneracy in the Kondo effect,
the effective field induced by the Kramers polarization enhances the
degeneracy through suppressing the spin-orbit coupling; ii) while the effective field induced by the
spin-polarization can not be compensated by applying magnetic field,
the effective field induced by the orbit-polarization can be
compensated; and iii) the presence of the
spin-orbit coupling does not change the compensation behavior observed in the case without the spin-orbit coupling. These results are observable in an ultraclean carbon-nanotube quantum dot attached to ferromagnetic contacts under a parallel applied magnetic field along the tube axis and it would deepen our understanding on the Kondo physics of the carbon nanotube
quantum dot.
\end{abstract}

\maketitle

\section{Introduction} \label{introd}

The experimental and theoretical studies of the Kondo effect \cite{Kondo1964} in
artificial confined systems has attracted much attention since 1998,
the first experimental observation of the Kondo effect in
semiconductor quantum dot. \cite{Gordon1998, Cronenwett1998, Schmid1998} The advantage of
the quantum dot as the platform studying the Kondo effect is its
tunability, namely, one can tune readily the voltages of various electrodes of the quantum dot to control the relevant model parameters in describing the Kondo effect, as a result, one can study in a deep way various aspects of the transport property in such a system.  For example, by tuning the gate
voltage one can observe the Coulomb blockade effect, \cite{Meirav1990, Meir1991} the Kondo effect \cite{Gordon1998, Cronenwett1998, Schmid1998} and its unitary limit, \cite{Wiel2000, Delft2000} and even from the Kondo regime to the mixed-valence regime. \cite{Gordon1998b} The non-equilibrium Kondo effect has also been studied by tuning the source-drain bias \cite{Simmel1999, Franceschi2002} and the couplings to the leads has been tuned to observe the offset of the Kondo resonance, \cite{Simmel1999} and so on.

Recently, due to the development of the spintronics the influence of
the polarized electrodes attached to the quantum dot has also been
intensively investigated experimentally \cite{Pasupathy2004, Hamaya2007, Sahoo2005, Hofstetter2010} and theoretically. \cite{Zhang2002, Sergueev2002, Martinek2003a, Martinek2003b, Sindel2007, Krawiec2007, Lim2010, Choi2004, Barnas2006} It was found that the effect of the polarized electrodes is equivalent to an effective exchange field and can be
compensated by applying an external magnetic field.
\cite{Pasupathy2004, Hamaya2007, Martinek2003a, Martinek2003b, Sindel2007, Krawiec2007, Lim2010}

The semiconductor quantum dot involved only one level is a simple system to study the Kondo
physics since in such a system only the spin degree of
freedom is involved. A slightly complicated system involved degree
of freedom other than the spin is the carbon nanotube (CNT) quantum
dot, which includes the orbital degree of freedom,
\cite{Ando2000, Nygard2000, Minot2004, Herrero2005, Choi2005, Makarovski2005} and the spin-orbit coupling could be stronger than the believed earlier, as predicted theoretically \cite{Hernando2006, Bulaev2008} and confirmed by the experiment. \cite{Kuemmeth2008} The presence of the spin-orbit coupling has a significant influence to the transport behaviors in the CNT quantum dot, for example, the $SU(4)$ symmetry discussed in the CNT quantum dot \cite{Herrero2005, Choi2005} is no longer valid and
correspondingly the Kondo resonance shows some interesting splitting
effects.\cite{Fang2008, Galpin2010} This motivates us to further study a question,
namely, how does the compensation effect observed in the
semiconductor quantum dot behave in the CNT quantum dot in the presence of strongly spin-orbit
coupling? In Ref. [\onlinecite{Lim2011}], the Kramers polarization due to the presence of the spin-orbit coupling and its influence on the Kondo peak splitting have been studied in detail by using scaling analysis and the slave-boson technique. \cite{Guillou1995, Krawiec2007} The compensation effect of the $SU(2)$ spin and the orbital Kondo effects were discussed in the absence of the spin-orbit coupling. In the present work, we study systematically the compensation effect with and without the spin-orbit coupling and the results show many novel features as presented later.

Experimentally, the spin polarized transport through CNT quantum
dots attached to ferromagnetic leads has been reported in the
literature. \cite{Hauptmann2008} When the external magnetic
field is applied perpendicularly, Hauptmann \textit{et. al.} found that the
exchange field can be compensated by the external field applied,
which is consistent with that observed in the semiconductor quantum
dot. The possible reason is that in the perpendicular case the spin
projection along the CNT axis is no longer a good quantum number and
as a result the response of the single-particle energy spectrum to
the applied field is approximately the Zeeman effect-like. \cite{Logan2009, Fang2010} This is
because that the perpendicular field only couples to the spin, not
to the orbital degrees of freedom. In the present work we focus on
the parallel magnetic field case, in which the presence of the
strong spin-orbital coupling should play a significant role in the
transport properties with the polarized electrodes.

There are three kinds of possible polarizations of conduction
electrons in CNT, namely, the spin-polarization, the orbit-polarization and the Kramers
polarization. \cite{Lim2011} In the absence of the spin-orbit coupling, the Kramers polarization is also absent. Thus we
discuss the effects of the spin- and orbit-polarizations and
compare them with those in the semiconductor quantum dots. In the
presence of the spin-orbit coupling, the Kramers polarization is
found to have a different feature in comparison to the spin- and orbit-polarizations. When both spin- and orbit-polarizations remove the degeneracy, the Kramers polarization can enhance the degeneracy. By applying magnetic field, the compensation behavior obtained is quite different to that found in the semiconductor quantum dot and in
the perpendicular field case. While the effective exchange field induced by the orbit-polarization can be compensated, the effective exchange field induced by the spin-polarization can not be compensated. Due to the interplay between the spin-orbit coupling, the polarized electrodes (spin-, orbit- and Kramers polarizations) and the magnetic field applied, the Kondo peaks
show complicated splitting behaviors. We analyze in detail the correspondences between these sub-peaks and their microscopic tunneling processes.

This paper is organized as follows. In Sec. \ref{section2} we use
the Anderson model to describe the CNT quantum dot and use the
Green's function formalism to study the polarized transport
behaviors. In Sec. \ref{section3} we present the explicit numerical
results and discuss microscopic tunneling processes. Finally, Sec.
\ref{section4} is devoted to a brief summary.

\section{Model and Green's function formalism} \label{section2}
The CNT quantum dot can be described by the Anderson impurity model \cite{Anderson1961}
\begin{eqnarray}
&& H = \sum_{km\alpha }\epsilon_{km\alpha }c_{km\alpha }^{\dagger}c_{km\alpha} + \sum_{m}\varepsilon _{m}d_{m}^{\dagger }d_{m} + \frac{U}{2}\sum_{m\neq m'}n_{m'}n_{m}  \notag \\
&&\hspace{3cm} +\sum_{km\alpha }\left( V_{\alpha }d_{m}^{\dagger}c_{km\alpha}+h.c.\right), \label{H}
\end{eqnarray}
where $d_{m}^{\dagger }(d_{m}) $ and $c_{km\alpha }^{\dagger
}(c_{km\alpha }) $ represent the creation (annihilation) operators
of an electron in the dot and the left ($\alpha = L$) and the right
($\alpha = R$) leads, respectively. Here $m = \{\sigma, \tau\}$
describes the configuration of electrons where $\sigma = \uparrow$
or $\downarrow$ and $\tau =\pm$ denote the spin and orbital quantum
numbers. $\epsilon_{km\alpha} $ is the single-particle energy
spectrum in the leads with the configuration $m$ and
$\varepsilon_{m}$ is the dot level related to the spin-orbit coupling, which will be given below. $n_m
= d_{m}^{\dagger}d_{m}$ is the occupation operator, $U$ is the
on-site interaction and $V_{\alpha}$ is the tunneling amplitude
between the dot and leads. Here we assume that the configuration $m$
of an electron is conserved during tunneling between the dot and
leads.

The electronic structure of the dot affects directly its transport properties obtained from the current through the dot within the framework of the
Keldysh formalism \cite{Meir1993, Jauho1994}
\begin{equation} \label{current}
I=\frac{i\,e}{\hbar }\frac{\Gamma_L \Gamma_R}{\Gamma_L + \Gamma_R}\sum_{m}\int\,d\omega \rho_{d,m}(\omega) \left( f_L(\omega) - f_R(\omega) \right),
\end{equation}
where $f_\alpha(\omega) $ is the Fermi distribution of the lead
$\alpha$ and $\Gamma_\alpha = \sum_m \Gamma_{\alpha, m}$ with
$\Gamma_{\alpha,m} = \pi \rho^0_{\alpha,m}|V_\alpha|^2$. Here
$\rho^0_{\alpha, m}$ denotes the density of states of the
polarized electrodes with the configuration $m$, which is related to
the spin polarization $P^s_\alpha$, the orbital polarization
$P^o_\alpha$ as well as the Kramers polarization $P^k_\alpha$ in the presence of strongly spin-orbit coupling. \cite{Lim2011} Thus the coupling matrix of different configurations can be expressed as follows
\begin{eqnarray} \label{pt}
&& \Gamma_{\alpha,\{\uparrow,+\}} = \frac{\Gamma_\alpha}{4}(1 + P^s_\alpha + P^o_\alpha + P^k_\alpha) \\
&& \Gamma_{\alpha,\{\uparrow,-\}} = \frac{\Gamma_\alpha}{4}(1 + P^s_\alpha - P^o_\alpha - P^k_\alpha) \\
&& \Gamma_{\alpha,\{\downarrow,+\}} = \frac{\Gamma_\alpha}{4}(1 - P^s_\alpha + P^o_\alpha - P^k_\alpha) \\
&& \Gamma_{\alpha,\{\downarrow,-\}} = \frac{\Gamma_\alpha}{4}(1 - P^s_\alpha - P^o_\alpha + P^k_\alpha)
\end{eqnarray}
In Eq. (\ref{current}) $\rho_{d, m} (\omega)=-\frac{1}{\pi}\text{Im}G^r_{d, m}(\omega)$ is the density of states
in the quantum dot, where the retarded Green's function
$G^r_{d,m}(\omega)$ can be obtained by using the equation
of motion approach, \cite{Zubarev1960} in which the hierarchy of equations has to be truncated at certain level. Here we consider the Lacroix approximation, \cite{Lacroix1981} which is enough to capture the compensation effect we are interested and the higher-order effects \cite{Luo1999} are neglected here.  Before the Green's
function $G^r_{d,m}(\omega)$ is derived we first discuss the quantum
dot level $\varepsilon_m$, which is given in the presence of the
parallel magnetic field \cite{Fang2008}
\begin{equation}
\varepsilon_m \equiv \varepsilon_{\{\sigma,\tau\}} = \varepsilon^{0}_{d}+
\frac12\sigma \tau \Delta_{so}+\tau \mu B+\sigma B,  \label{epsm}
\end{equation}
where $\varepsilon^{0}_{d}$ is bared level depending on the
geometric parameters of the dot and the gate voltage. The second
term is due to the spin-orbit coupling. \cite{Hernando2006, Kuemmeth2008, Bulaev2008, Galpin2010}
The third term is the orbital Zeeman splitting, where $B=g\mu_{B}\mathbf{B}$ is
the renomalized magnetic field, $\mu =2\mu _{orb}/\left( g\mu
_{B}\right) $ is the ratio between the orbital magnetic moment $\mu _{orb}$ and
the Bohr magneton $\mu _{B}$ and $g$ is the Land\'e g-factor. It was found that $\mu_{orb}$ is usually $10\sim 20$ times
larger than $\mu_B$. \cite{Minot2004, Kuemmeth2008} The last term is the spin zeeman splitting in the presence of magnetic fields.

Due to the presence of the polarized electrodes, the dot level
(\ref{epsm}) would be further modified. In the semiconductor quantum
dots, the ferromagnetic electrodes induce an effective exchange
field due to the spin-dependent charge fluctuations.
\cite{Martinek2003a, Martinek2003b, Choi2004, Pasupathy2004, Sindel2007} This
behavior can be well understood by Haldane's scaling theory. \cite{Haldane1978} In the CNT quantum
dots the orbit degree of freedom comes into play. The spin-, orbit-polarizations as well as possible Kramers-polarization in the presence the spin-orbit coupling can also induce effective exchange fields on the dot levels. By the same way one can obtain modified dot levels
$\tilde{\varepsilon}_{m}=\varepsilon_{m}+\delta \varepsilon_{m }
(\text{note}\, m = \{\sigma,\tau\})$, where
\begin{eqnarray}
&& \delta \varepsilon_{\{\sigma,\tau\}}=\sum_{\alpha }\int
\frac{d \varepsilon }{\pi }\left( \frac{\Gamma_{\alpha,\sigma\tau }( 1-f_{\alpha }(\varepsilon) )
}{\varepsilon_{\sigma\tau}-\varepsilon }+\frac{\Gamma
_{\alpha,\sigma\bar{\tau} }f_{\alpha }(\varepsilon)
}{\varepsilon -U-\varepsilon _{\sigma \bar{\tau}}} \right.\nonumber \\
&& \hspace{2.5cm} \left.+\frac{ \Gamma_{\alpha,\bar{\sigma}\tau }f_{\alpha }(\varepsilon)
}{ \varepsilon -U-\varepsilon _{\bar{\sigma}\tau }}+\frac{\Gamma
_{\alpha,\bar{\sigma}\bar{\tau}}f_{\alpha }(\varepsilon)}{\varepsilon -U-\varepsilon
_{\bar{\sigma}\bar{\tau}}}\right).  \label{mod-level}
\end{eqnarray}
Here $\bar\sigma(\bar\tau) = -\sigma(\tau)$.
The first term in Eq. (\ref{mod-level}) corresponds to the charge
fluctuations between a single occupied state and empty
state in the dot levels. The remaining terms reflect
the charge fluctuations between the single occupied state and double occupied states with on-site Coulomb repulsion $U$. In the semiconductor quantum dot, the modified dot levels can be attributed to an effective exchange field and the field can be compensated by external magnetic field applied. \cite{Martinek2003a, Martinek2003b, Choi2004, Pasupathy2004, Sindel2007} Here we explore the possible compensation effects in the CNT quantum dot.

With the effective dot levels at hand, one can derive the Green's function $G^r_{d,m}(\omega)$, which reads
\begin{widetext}
\begin{equation}
G^r_{d,m}(\omega) = \frac{1-\sum_{m' \neq m}\langle n_{m'}\rangle -\sum_{m' \neq m}A_{mm'}(\omega)}{\omega - \tilde\varepsilon_m - \Delta_m(\omega) + \Delta_m(\omega)\sum_{m'\neq m}A_{mm'}(\omega)-\sum_{m'\neq m}B_{mm'}(\omega)}. \label{gd1}
\end{equation}
\end{widetext}
Here the Lacroix's approximation \cite{Lacroix1981} is used and for simplify, we also
consider the limit of $U\rightarrow\infty$. In this case, only the first term survives in Eq. (\ref{mod-level}). In Eq. (\ref{gd1}), $\langle n_{m}\rangle = \int \rho_{d,m}(\omega) f_m(\omega)
d\omega$ is the average occupation number of the configuration $m$
in the dot with $f_m(\omega) =
\frac{1}{\Gamma_m}\sum_\alpha\Gamma_{\alpha,m}f_\alpha(\omega)$ and
$\Gamma_m = \sum_\alpha \Gamma_{\alpha,m}$. The other notations
introduced are
\begin{eqnarray}
&& \Delta_m(\omega) = \sum_{k\alpha}\frac{|V_\alpha|^2}{\omega - \epsilon_{km\alpha}},\label{delta0}\\
&& A_{mm'}(\omega) = \sum_{k\alpha }\frac{V_{\alpha}\langle d_{m'}^{\dagger }c_{km'\alpha }\rangle }{\omega -\tilde\varepsilon_{m} + \tilde\varepsilon_{m'}-\epsilon _{km'\alpha }}, \label{amm} \\
&& B_{mm'}(\omega) = \sum_{kk'\alpha\alpha'}\frac{V_{\alpha }V_{\alpha'}^{\ast }\langle c_{k'm'\alpha'}^{\dagger}c_{km'\alpha }\rangle }{\omega -\tilde\varepsilon_{m}+\tilde\varepsilon _{m'}-\epsilon _{km'\alpha }}. \label{bmm}
\end{eqnarray}
By performing the summations of $k$ in Eqs. (\ref{delta0}), one has
\begin{equation}
\Delta_m(\omega) = \frac{\Gamma_m}{\pi} \int d\omega' P\frac{1}{\omega - \omega'} - i\Gamma_m \approx -i\Gamma_m \label{deltam},
\end{equation}
where ``P" denotes the principal part integration and the last
approximation is obtained by neglecting the real part of
$\Delta_m(\omega)$ for simplify. Similarly, $A_{mm'}(\omega)$ and
$B_{mm'}(\omega)$ can be written as
\begin{eqnarray}
&& A_{mm'}(\omega) = \frac{\Gamma_{m'}}{\pi }\int d\omega'\frac{f_{m'}(\omega')\left( G_{d,m'}^{r}(\omega') \right)^{\ast }}{\omega +i\eta -\tilde\varepsilon_{m}+\tilde\varepsilon _{m'}-\omega'}, \label{amm-2}\\
&& B_{mm'}(\omega) = \frac{\Gamma_{m'}}{\pi} \int d\omega' \frac{f_{m'}(\omega')}{\omega +i\eta -\tilde\varepsilon_{m}+\tilde\varepsilon _{m'}-\omega'} \nonumber \\
&& \hspace{1cm} + \frac{i\tilde \Gamma_{m'}}{\pi} \int d\omega' \frac{\tilde f_{m'}(\omega') \left(G_{d,m'}^{r}(\omega')\right)^\ast}{\omega +i\eta -\tilde\varepsilon_{m}+\tilde\varepsilon _{m'}-\omega'}, \label{bmm-2}
\end{eqnarray}
where $\tilde{\Gamma}_{m'} = 2\Gamma_{L,m'}\Gamma_{R,m'}$ and
$\tilde f_{m'}(\omega) = \frac{1}{\tilde \Gamma_{m'}} \sum_{\alpha}
\Gamma_{\alpha,m'}\Gamma_{\bar\alpha,m'} f_{\alpha}(\omega)
(1-f_{\bar\alpha}(\omega))$ and here $\alpha = (L,R)$ and
$\bar\alpha = (R,L)$. As a result, the Green's function
$G^r_{d,m}(\omega)$ can be calculated self-consistently. It is well known that while the Lacroix's decoupling can capture correctly the Kondo effect, the height of the Kondo peak can not be obtained accurately. The reason is that in this scheme the occupation number can not be calculated accurately. However, in the present work we focus on the compensation effect, which is only related to the positions of the poles of the dot Green's function. From Eqs. (\ref{gd1}), (\ref{amm-2}) and (\ref{bmm-2}) one notices that the inaccuracy of the occupation number $\delta n$ will lead to an error of the position of the poles in the order of $\delta n/\varepsilon^0_d$, which can be safely neglected in the Kondo regime we are interested in.

\section{Results and discussion}\label{section3}
In this section, we discuss in detail the various polarization
effects by solving numerically the dot Green's function Eq.
(\ref{gd1}). While the interesting physics is involved only in the
parallel configuration of polarization, the spin- and
orbit-dependent charge fluctuations in the anti-parallel
configuration cancel out each other due to the anti-orientation
polarization of the leads. \cite{Martinek2003a, Martinek2003b, Choi2004, Pasupathy2004} Therefore, we only
consider the parallel configuration. In addition, since we are interested in the compensation behaviors of the spin- and orbit-polarizations as well as the Kramers polarization in the presence of the spin-orbit coupling, we neglect the possible correlation between the different polarizations \cite{Lim2011} and allow independent change of different polarizations. In the following calculations
we take $\Gamma_L = \Gamma_R = \Gamma$ as the units of energy and
fix $\varepsilon^0_d = -7.5\Gamma$.

\begin{figure}[hbp]
\includegraphics[width=\columnwidth]{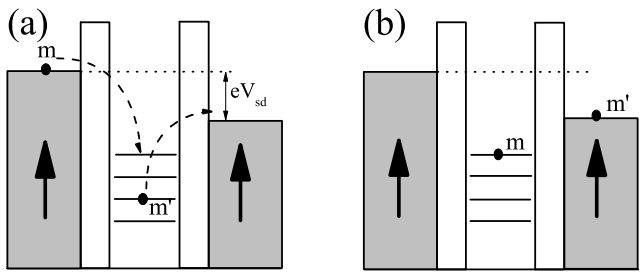}
\caption{Schematic diagram of the cotunneling process. (a) The configuration before the tunneling and (b) the configuration after the tunneling.  There
are six possible tunneling processes involving different spin and orbit flips in carbon nanotube quantum dot ($m \Leftrightarrow m'$).}
\label{fig1}
\end{figure}
Due to the presence of the spin and orbit degrees of freedom in the
system studied, the Kondo resonances can originate from different
cotunneling processes, as shown in Fig. \ref{fig1}.
After some careful analysis, it is found that there are twelve possible
cotunneling processes. As a result, the position of the Kondo
resonance obtained from each individual process can be determined by
the energy difference between the final and initial configurations,
namely, $eV_{sd} = \tilde\varepsilon_{m'} -
\tilde\varepsilon_{m}$. For convenience, we mark these different
contunneling processes by numbers. Explicitly, we use ``1" and ``4"
to denote the tunnelings with both spin and orbit flip,  ($|\uparrow+\rangle
\leftrightarrow |\downarrow-\rangle $) and ($|\uparrow-\rangle \leftrightarrow
|\downarrow+\rangle$), respectively and ``2" and ``5" to represent the tunnelings with only spin-flip and the orbit remains, i.e., ($| \uparrow+\rangle
\leftrightarrow |\downarrow+\rangle $ ) and ($|\uparrow-\rangle \leftrightarrow
|\downarrow-\rangle$), respectively and finally ``3" and ``6" mean
($|\uparrow+\rangle \leftrightarrow |\uparrow-\rangle$) and ($|\downarrow-\rangle
\leftrightarrow |\downarrow+\rangle$), respectively, namely, the spin keeps unchanged but the orbit flips. These six cotunneling
processes will lead to twelve Kondo peaks at most when all
degeneracy is lifted by interactions involving spin and orbit
degrees of freedom. In the following we explicitly discuss various
cases. We study possible compensation effects by applying magnetic
field and compare them with the results in the $SU(2)$ case.
\cite{Martinek2003a, Martinek2003b, Pasupathy2004, Sindel2007, Hamaya2007, Krawiec2007, Lim2010} For
completeness, we firstly discuss the case that the spin-orbit
coupling is neglected.

\subsection{Polarization effects without the spin-orbit coupling}
\subsubsection{Spin-polarization effect}
\begin{figure}[hbp]
\includegraphics[width=\columnwidth]{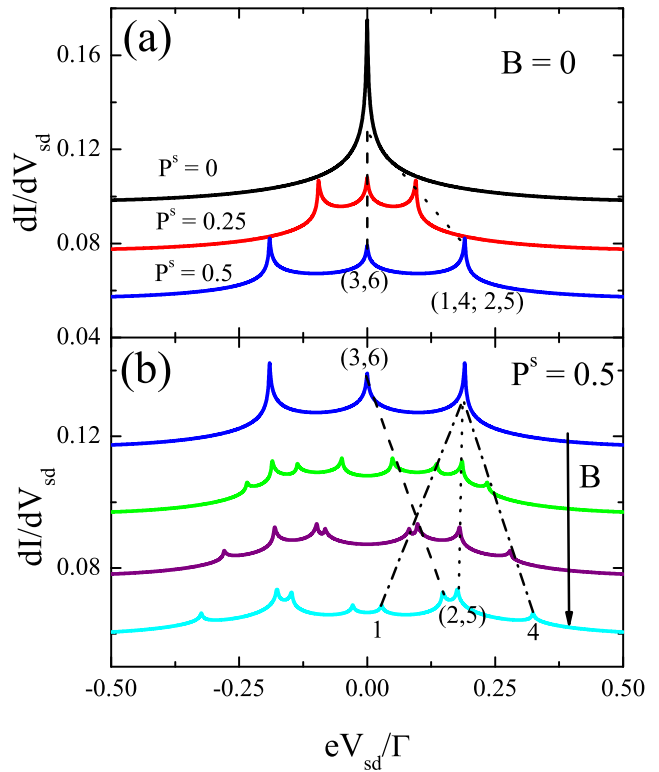}
\caption{The spin-polarization effect and its compensation behavior in the absence of the spin-orbit coupling. (a) The Kondo peak splitting due to the
spin-polarization and (b) the Kondo sub-peaks evolution under the external magnetic field applied. For clarity, here and hereafter the curves have been shifted perpendicularly to show the change of the sub-peak position with the polarization in (a) and the field in (b). The parameters used are $P^o = P^k = 0$ and $\Delta_{so} = 0$.  In (b) the magnetic fields applied are $0, 0.0025\Gamma, 0.005\Gamma$ and $0.0075\Gamma$ (from top to bottom).} \label{fig2}
\end{figure}

In this case, the Kondo physics in the CNT quantum dot
involves spin and orbit degrees of freedom but they are independent
of each other, thus the system has the $SU(4)$ symmetry. \cite{Choi2004}
There are two polarization effects to be considered, namely, spin-
and orbit-polarizations. In Fig. \ref{fig2} we firstly show the
spin-polarization effect and its possible compensation by applying external
magnetic field. When there is no any polarization in the
electrodes, all energy levels in the dot are degenerate, so the Kondo
resonance shows a single peak, as shown in Fig. \ref{fig2} (a) for
$P^s = 0$. When the spin polarization is switched on, the Kondo
peak splits into three sub-peaks, where the processes ``3" and ``6"
are unaffected by the spin-polarization but the remaining processes
are related to the spin-polarization. This is because that the
former does not involve the spin flip but the other processes do
involve the spin flip. Moreover, the more stronger the
spin-polarization, the more larger is the distance between these
three sub-peaks. Furthermore, these three sub-peaks can not be
compensated by applying external magnetic field, as shown in Fig. \ref{fig2} (b). This is in contrast to the statement that the spin-polarization can be compensated by applying an external magnetic field. \cite{Lim2011}
On the contrary, with increasing magnetic field, these
three sub-peaks split further due to orbit- and spin-Zeeman
effects.  It is noted that the processes ``2" and ``5" do not change
with increasing magnetic field since in these two processes the
orbital degree of freedom is not changed. Below we consider the
orbit polarization effect.

\subsubsection{Orbit-polarization effect}
\begin{figure}[hbp]
\includegraphics[width=\columnwidth]{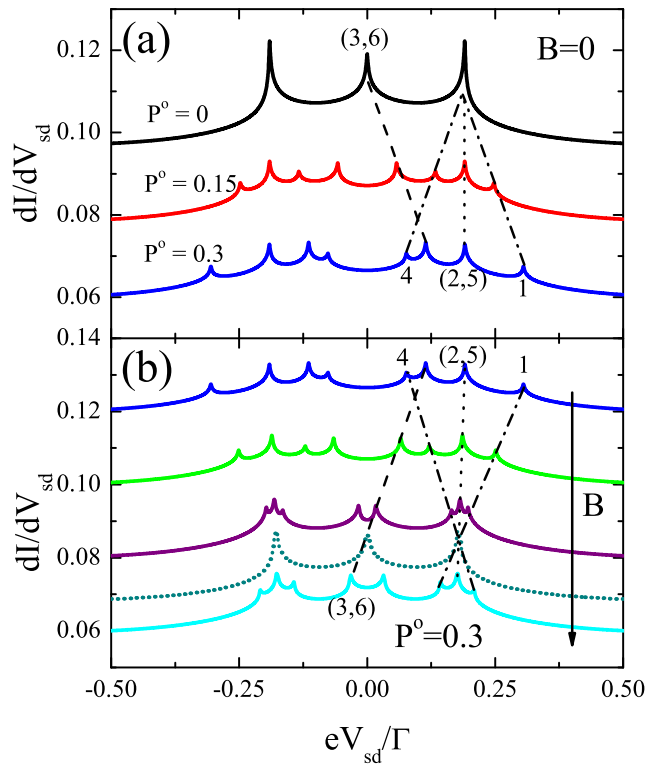}
\caption{The orbit-polarization effect and its compensation behavior. (a) The Kondo sub-peaks splitting due to the orbit polarization and (b) the Kondo sub-peaks evolution under the external magnetic field applied. The parameters used are $P^k = 0$ and $\Delta_{so} = 0$ and $P^s = 0.5$. In (b) the magnetic fields applied are $0, 0.0025\Gamma, 0.005\Gamma, 0.006\Gamma, 0.0075\Gamma$ (from top to bottom). The dotted line denotes the complete compensation effect at $B = 0.006\Gamma$.} \label{fig3}
\end{figure}

When the orbit-polarization is switched on, the orbital degeneracy
is removed. As a result, Fig. \ref{fig3} (a) shows how the Kondo
peaks split with increasing of the orbit polarization. It is observed that the processes ``2" and ``5" do not change with the change of the orbit
polarization but the degeneracy between the processes ``1" and ``4" is removed. To compare Fig. \ref{fig3}(a) to Fig. \ref{fig2} (b), the evolution of the Kondo sub-peaks under the external magnetic field without the orbit polarization is the same as that with the orbit polarization with zero field. This means that the effective exchange field induced by the orbit polarization can be completely compensated by the external magnetic field. This is true, as shown in Fig. \ref{fig3} (b) (dash-dotted line). With increasing field, the processes ``1" and ``4" separated by the orbit polarization merge again at $B = 0.006\Gamma$. This is in contrast to the spin-polarization case. The compensation effect
of the orbit polarization observed here has been not reported in the
literature. The physical reason of this compensation effect is that
the orbit Zeeman effect dominates in such a system and the spin
Zeeman is very weak. In the following we consider the case that the
spin-orbit coupling is present.
\begin{figure}[hbp]
\includegraphics[width=\columnwidth]{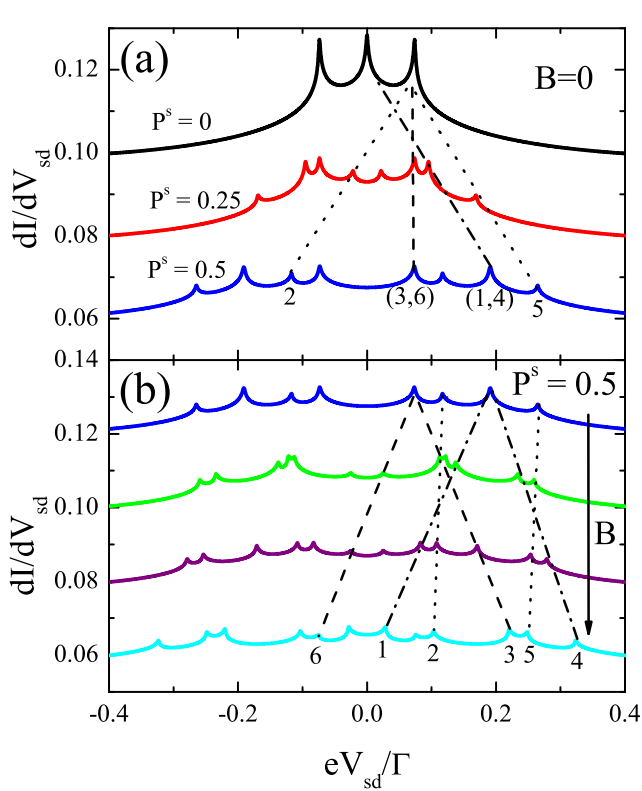}
\caption{The spin-polarization effect and its compensation behavior in the presence of the spin-orbit coupling. (a) The Kondo peak splitting due to the
spin-polarization at zero field and (b) the Kondo sub-peaks evolution under the external magnetic field applied when $P^s = 0.5$. The parameters used are $P^o = P^k = 0$ and $\Delta_{so} = 0.0075\Gamma$. In (b) the magnetic fields applied are $0, 0.0025\Gamma, 0.005\Gamma, 0.0075\Gamma$ (from top to bottom).} \label{fig4}
\end{figure}
\begin{figure}[hbp]
\includegraphics[width=\columnwidth]{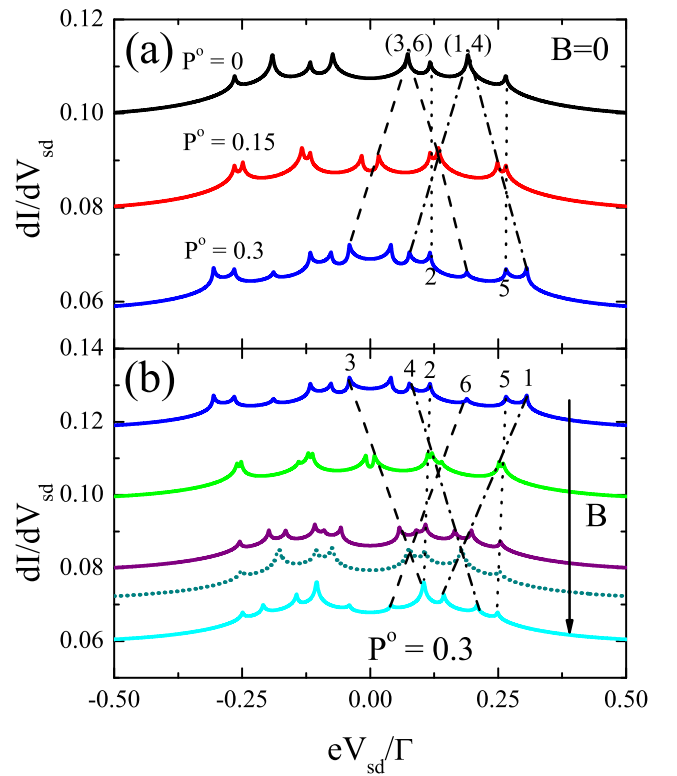}
\caption{The orbit-polarization effect in the presence of the spin-orbit coupling and its compensation behavior. (a) The Kondo sub-peaks splitting due to the orbit polarization at zero field and (b) the Kondo sub-peaks evolution under the external magnetic fields applied when $P^o = 0.3$. The parameters used are $P^k = 0$ and $\Delta_{so} = 0.0075\Gamma$. In (b) the magnetic fields applied are $0, 0.0025\Gamma, 0.005\Gamma, 0.006\Gamma, 0.0075\Gamma$ (from top to bottom). The dotted line denotes the compensation effect at $B = 0.006\Gamma$.} \label{fig5}
\end{figure}
\begin{figure}[hbp]
\center{\includegraphics[width=\columnwidth]{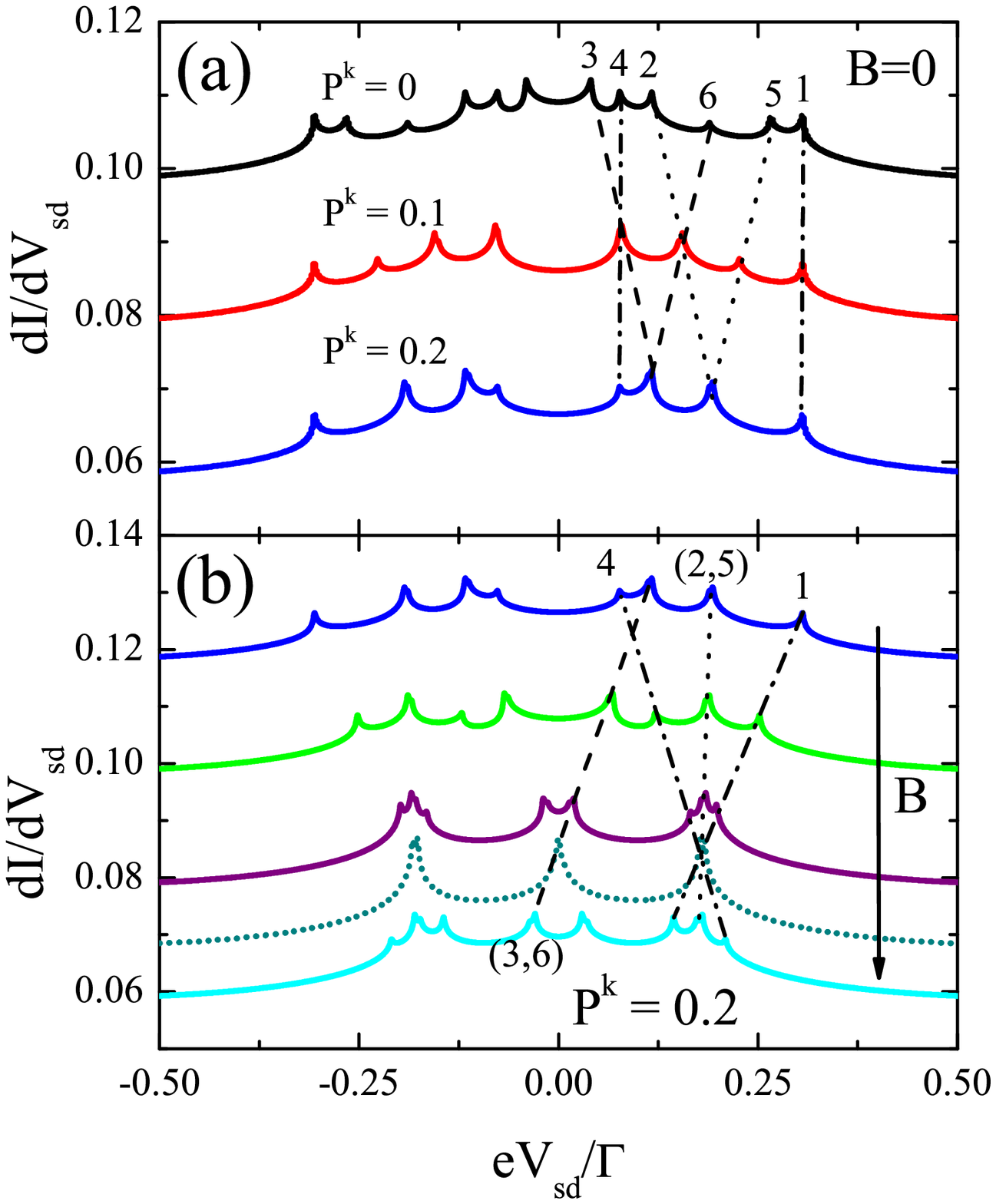}}
\caption{The Kramers polarization effect in the presence of the spin-orbit coupling and its compensation behavior. (a) The Kondo sub-peaks merging due to the Kramers polarization at zero field and (b) the Kond sub-peaks evolution under the external magnetic fields applied with $P^k = 0.2$. The parameter used are $\Delta_{so} = 0.0075\Gamma$. In (b) the magnetic fields applied are $0, 0.0025\Gamma, 0.005\Gamma, 0.006\Gamma, 0.0075\Gamma$ (from top to bottom). The dotted line denotes the compensation effect at $B = 0.006\Gamma$.} \label{fig6}
\end{figure}

\subsection{Polarization effect with the spin-orbit coupling}
In the presence of spin-orbit coupling, the spin- and orbit-degrees
of freedom are no longer independent of each other and as a result,
the SU(4) Kondo physics breaks down. \cite{Fang2008} In this case, even there is no
magnetic field, the Kondo resonance shows three sub-peaks, as shown
in Fig. \ref{fig4}(a) for $P^s = 0$. The peak located at the
center is due to the processes ``1" and ``4", the side-peaks
originate from the remaining processes, as pointed out in the
previous work. \cite{Fang2008} This is in contrast to the
single peak observed in Fig. \ref{fig2}(a). Below we consider the
polarization effects in the electrodes. As in the last section, we
firstly consider the spin-polarization effect.

\subsubsection{Spin-polarization effect}
With increasing the spin-polarization, the processes ``3" and ``6"
keep unchanged, which is the same as Fig. \ref{fig2} without the
spin-orbit coupling. The reason is the same there. However, the
remaining processes is completely different. While the degeneracy
between (1,4) and (2,5) is removed by the spin-orbit coupling,
the processes (1,4) keep untouched by the spin-polarization, the
processes (2,5) are sensitive to the spin-polarization. This is
because that in the processes (1,4) the effective field induced by
the spin-polarization does not change the symmetry of the spin and
orbit since the orbit flips together with the spin. This is not
true for the processes (2,5) in which the orbital keeps unchanged
when the spin flips. As a result, the degeneracy between the
processes ``2" and ``5" is quickly removed with increasing the
spin-polarization. Interestingly, by applying the magnetic field,
the processes (2,5) keep unchanged, as shown in Fig. \ref{fig4}(b),
the degenerates between the processes ``1" and ``4" and the
processes ``3" and ``6" are removed. In this case, all degeneracy in
such a system has been lifted and the Kondo resonance shows twelve
sub-peak structure, as observed in Fig. \ref{fig4}(b). Here the
magnetic field does not play a compensation role, on the contrary,
it removes all possible degeneracy.

\subsubsection{Orbit-polarization effect}
Now we discuss the orbit-polarization effect in the presence of
spin-orbit coupling. When increasing the orbit polarization, one
notes that the processes ``2" and ``5" do not change since in these
two processes the orbital degree of freedom does not flip. However,
the degeneracy between the processes ``1" and ``4" is lifted. The
processes ``3" and ``6" so do. As a consequence, all degeneracy in
such a system is removed by the presence of spin- and
orbit-polarizations. This can be seen from Fig. \ref{fig5} (a), in
which there are twelve sub-peaks for $P^o = 0.3$. By applying
magnetic field, one finds that the splittings of (1,4) and (3,6) observed in Fig. \ref{fig5}(a) can be partly removed,
as shown by the dotted line in Fig. \ref{fig5}(b), which is roughly
consistent with the curve of $P^o = 0$ in Fig. \ref{fig5}(a).
This is exactly the compensation effect induced by the orbit
polarization.

\subsubsection{Kramers-polarization effect}
In the presence of the spin-orbit coupling, a novel polarization
effect can be proposed, namely, the Kramers polarization. As
discussed above, the presences of the spin- and orbit-polarizations
can remove all degeneracy in carbon nanotube quantum dot and the
Kondo resonance can show all possible twelve sub-peak structure.
However, the additional Kramers polarization plays an inverse role,
namely, it leads to the degeneracy in such a system, as shown in
Fig. \ref{fig6} (a). While the processes (1,4) are not be affected
by the Kramers polarization since in such processes the spin and
orbit flip together, the non-degeneracy of the processes ``2" and
``5" is removed when $P^k=0.2$. The same effect is also
seen in the processes ``3" and ``6". Therefore, the Kramers
polarization plays an effective field to remove the non-degeneracy
leaded by the spin-orbit coupling. This effective field can not be compensated
by applying magnetic field, as observed in Fig. \ref{fig6}(b). With
increasing magnetic field, the processes (2,5) keep untouched, and
remaining processes are found to depend strongly on the applied
magnetic field. Physically, in the processes (2,5) the orbit does
not flip but in the remaining processes the orbit degree of freedom
flips. The three peaks structure observed in Fig. \ref{fig6}(b)
(the dotted line) corresponds to the spin-polarization splitting of the
$SU(4)$ Kondo peak since the spin-orbit coupling is suppressed
completely by the Kramers polarization and the effective field induced by the orbit polarization
are compensated by the external magnetic fields. This indicates that the effective field induced by the orbital polarization can always be compensated by
applying external magnetic field, which can also be seen from Fig. \ref{fig6}(b), in spite of the presence of the spin-orbit coupling.

\section{Summary}\label{section4}
In this paper we study the compensation effect in CNT
quantum dot involving spin and orbit degrees of freedom. In
comparison to the semiconductor quantum dot, the polarized
electrodes attached to the dot contain three polarization effects,
namely, the spin- and orbit-polarization as well as the Kramers
polarization in the presence of the spin-orbit coupling. In the
semiconductor quantum dot, the effective field induced by the
spin-polarization can be completely compensated by applying magnetic field.
However, here in the CNT quantum dot, one finds that
only the effective field induced by the orbit polarization can be
compensated by applying magnetic field and that induced by the
spin-polarization can not be compensated. This can be due to that
the orbital magnetic moment is much larger than the spin magnetic
moment. This is different from the statement that the $SU(2)$ spin Kondo effects can be compensated by applying an external magnetic field. \cite{Lim2011}
In addition, we also find that the effective fields induced
by the spin- and orbit-polarizations remove the degeneracy but the
effective field induced by the Kramers polarization can enhance the
degeneracy through suppressing the spin-orbit coupling.  Furthermore, the
presence of the spin-orbit coupling does not change the compensation
effect on the effective field on the orbit degree of freedom induced
by any polarization. In such a system, the compensation effect of
the effective field induced by the orbit polarization has not been reported in the literature
and could be tested in the future experiments.

\begin{acknowledgments}
Support from CMMM of Lanzhou University, the NSF-China, the national program for basic research and the
fundamental research funds for the central universities of China is acknowledged.
\end{acknowledgments}

\end{document}